\documentclass{caosp310}
\usepackage{graphicx}
\usepackage{natbib}
\usepackage{float}
\articleNo{375}
\pubyear{2026}
\volume{56}
\volnumber{3}
\firstpage{1}
\received{February 19, 2026}
\accepted{March 27, 2026}

\begin{document}
\htitle{Shape model and dynamical state of the asteroid (300) Geraldina}
\hauthor{G.\,Apostolovska, N.\,Todorovi\'{c}, E.\,Vchkova Bebekovska, G.\,Borisov, A.\,Kostov and Z.\,Donchev}
\title{Shape model and dynamical state of the asteroid (300) Geraldina: Implications for its possible ancient origin}

\author{
        G.\,Apostolovska\inst{1}\orcid{0000-0002-5355-431X}
      \and
        N.\,Todorovi\'{c}\inst{2}\orcid{0000-0003-4475-750X}
      \and 
        E.\,Vchkova~Bebekovska\inst{1}\orcid{0000-0002-4872-7378}
      \and 
        G.\,Borisov\inst{3}\orcid{0000-0002-4516-459X}
			\and
        A.\,Kostov\inst{3}\orcid{0000-0003-0569-4542}
			\and
        Z.\,Donchev\inst{3}
       }

\institute{
           Institute of Physics, Faculty of Science, Ss. Cyril and Methodius University in Skopje,  Arhimedova 3, 1000 Skopje, Republic of Macedonia \email{gordanaapostolovska@gmail.com}
         \and 
           Astronomical Observatory, Volgina 7. P.O. Box 74, 11060 Belgrade, Serbia
         \and 
           Institute of Astronomy and National Astronomical Observatory,  Bulgarian Academy of Sciences, Tsarigradsko Chaussee Blvd. 72, BG-1784, Sofia, Bulgaria
          }

\date{February 19, 2026}

\maketitle

\begin{abstract}
In this study, we present a comprehensive photometric and physical characterization of the main-belt asteroid (300) Geraldina. Our analysis includes determining its sidereal rotational period, shape modeling, spin-axis orientation, as well as dynamical and spectral properties. The investigation is based on two decades of archival photometry from the Bulgarian National Astronomical Observatory (BNAO) Rozhen, complemented by dense CCD lightcurve observations obtained since 2017 at the Astronomical Station Vidojevica (ASV), dense data from ALCDEF, and augmented with sparse-in-time measurements from Gaia Data Release 3 (DR3). By combining these heterogeneous datasets within and using the lightcurve inversion method, we confirmed the asteroid’s prograde sense of rotation and obtained two symmetrically mirrored pole solutions for the asteroid model. Our dynamical studies show that (300) Geraldina remains stable in 100\,Myr. Although close to the 2:1 mean-motion resonance with Jupiter, it is not captured into it during the observed time. Instead, it exhibits interactions with an unidentified high-order mean-motion resonance, which appears to contribute to its long-term stability. Its spectral type (C) with a primitive carbonaceous composition, in combination with low albedo and long-term stability, is consistent with the assumption that (300) Geraldina is an ancient asteroid.
\end{abstract}

\section{Introduction}
\label{intr}

Determining the asteroid model is an important step for further investigation of the evolution, dynamics, and future of our Solar System. Although the number of asteroids with known models is steadily increasing, it still represents only a small fraction of the total known asteroid population (about 0.8\%). Specifically, out of 1,298,026 objects listed in the AstDyS\footnote{https://newton.spacedys.com/astdys2/index.php?pc=1.0.0} database, only 10,758 have shape models available in the DAMIT\footnote{https://damit.cuni.cz/} database (in December 2025). Having an asteroid with a known model means that both synodic and siderial periods, as well as the coordinates of its axis of rotation, the sense of rotation, and its shape, are known.

Most asteroids observed today are believed to be fragments of larger parent bodies that were disrupted in past collisional events. These fragments typically remain near the original parent body until dynamical processes gradually drift them away. Such groups of related objects are known as asteroid families, in which the largest member is usually considered the remnant of the original body. Studying asteroid models helps in the reconstruction of these collisional events, and also improves our understanding of the dynamical evolution of the Solar System as a whole.
By 2015, a total of 122 asteroid families had been confirmed \citep{2015aste.book..297N}, and this number remained relatively unchanged until 2024, when \cite{2024ApJS..274...25N} discovered more than 150 new collisional main belt families. Combined with Trojans and other families outside the main belt, the current number of known asteroid families exceeds 300. However, only three families are estimated to have an age of more than 3\,Gyr, originating from ancient asteroids.

The asteroid (300) Geraldina, discovered in 1890 by Auguste Charlois, is an outer main-belt asteroid and is considered a possible member of the so-called ancient asteroid population. These bodies are regarded as primordial remnants of the early Solar System and are therefore of particular interest for studies of the collisional and dynamical history of the main belt. The reported diameter in the MP3C database\footnote{Available at https://mp3c.oca.eu/}, varies from 66.78\,km \citep{2012ApJ...759L...8M} to 91.56\,km \citep{2016AJ....152...63N}, whereas the reported albedo ranges  from 0.03  \citep{2016AJ....152...63N} to 0.06 \citep{2014ApJ...791..121M}. A relatively large size, lack of association with any known asteroid family, and a low albedo, make (300) Geraldina a strong candidate for a pristine main-belt object - an ancient asteroid \citep{2017Sci...357.1026D, 2023A&A...676A...5F, 2022A&A...666A.116A, 2024A&A...690A.215A}.

According to the first studies on the rotation of this asteroid, it is derived that its synodic period is 6.818\,h \citep{2002ESASP.500..505I}. Later, a few more studies have been conducted on the ligthcurve of asteroid (300) Geraldina, and all of them show similar values for its rotational period. The latest value given in the JPL Database is 6.8423\,h. \cite{2018MPBu...45..162K} suggested a possible bimodal period of 13.726\,h.

Currently, the model of this asteroid is unknown. The only suggestion so far is given by \cite{2024MPBu...51..213R}, indicating the possibility of a binary asteroid, specifically a contact binary or an elongated single object. Still, this suggestion was based on three observational nights during one opposition, which can not provide sufficient data points for a clear statement. To construct the asteroid’s model, we need photometric observations from several different oppositions, to collect data from various parts of the asteroid’s surface and piece them together like a puzzle to form a picture; typically, this requires more than five observing nights, over a longer time period, and much more if the asteroid has an irregular shape. To decrease the observational time on the ground-based telescopes, photometric dense data can be combined with sparse data from space missions like NEOWISE or GAIA. While dense data from a single observational night allows us to collect hundreds of photometric points \citep{2019AIPC.2075i0016B, 2023ibpu.confE..39E, 2023ibpu.confE..41A}, sparse data are rare in time and typically provide only a few points (usually 5) per observational night, but can cover parts of the asteroid that have not been observed during the ground-based observations.

\section{Instrumentation}

Observations of (300) Geraldina were performed during a time span of 23 years. In this period, five different telescopes located in two observatories were used. The campaign began at the Bulgarian National Astronomical Observatory (BNAO) Rozhen, Bulgaria, with the 2m Ritchey-Chr\'{e}tien-Coude (RCC) telescope, the main instrument of the observatory. The telescope is equatorial mounted, and has two optical systems: Ritchey-Chr\'{e}tien (for optical observations) and Coude (for spectroscopic observations). The main mirror is two meters in diameter, the focal length is 16\,m. The second telescope of BNAO is the 50/70cm Schmidt Telescope - equatorial mounted Schmidt system with 70\,cm mirror and 50\,cm diameter of the correction plate, and a focal length of 172\,cm. The third is the 60cm Cassegrain telescope - equatorial mounted Cassegrain optical system with a main mirror of 60\,cm in diameter and a focal length of 7.5\,m. 

The Astronomical Station Vidojevica (ASV), Serbia, participated with two more telescopes. The first one is the 1.4m Milankovi\'{c} telescope - alt-azimuthal mounted Ritchey-Chr\'{e}tien system with main mirror of 1.4\,m in diameter, and focal length of 11.2\,m. The second instrument of the observatory is the 60cm Cassegrain Nedeljkovi\'{c} telescope - equatorial mounted Cassegrain optical system with main mirror of 60\,cm in diameter, and focal length of 6\,m.

The detectors of the telescopes, the component of the observational system that is most frequently updated, were also changed during the observational campaign, leading to 8 different CCD's used in total. In Tab.\,\ref{table_instruments} all these CCD's are listed. In the first column is the observational run (or period) in which the detector was used, followed by the telescope, where that CCD was attached, chip size in pixels, and the physical dimension of the pixel in $\mu$m.

\begin{table}
\small
\begin{center}
\caption{Instrumentation used in the observational campaign}
\label{table_instruments}
\begin{tabular}{lllcc}
Run	&	Telescope	&	CCD Detector Model	&	Chip size		&	Pixel size\\
		&						&											&		[px]			&	[$\mu$m]\\
\hline
\hline
2002 Jan	&	2mRCC	&	Photometrics CE200A-SITe	&	1024 x 1024	&	24\\
2003 Apr	&	5070	&	SBIG ST-8E								&	1520 x 1020	&	9\\
2008 Jan	&	5070	&	SBIG STL-11000M						&	4008 x 2672	&	9\\
2009 $\div$ 2010	&	5070	&	FLI PL16803	&	4096 x 4096	&	9\\
2011 $\div$ 2023	&	60CAS	&	FLI PL09000	&	3056 x 3056	&	12\\
2017 Oct	&	EQ60	&	Apogee U42	&	2048 x 2048	&	13.5\\
2018 $\div$ 2025	&	AZ1400	&	Andor iKon-L 936	&	2048 x 2048	&	13.5\\
2023 Oct	&	EQ60	&	FLI PL23042 MB	&	2048 x 2048	&	15\\
\hline
\hline
\end{tabular}
\end{center}
Notes: The telescope abbreviations are as follows:  2mRCC = 2m Ritchey-Chr\'{e}tien-Coude (RCC) telescope of BNAO Rozhen; 5070 = 50/70cm Schmidt telescope of BNAO Rozhen; 60CAS = 60cm Cassegrain telescope of BNAO Rozhen; EQ60 = 60cm Cassegrain Nedeljkovi\'{c} telescope of AS Vidojevica; AZ1400 = 1.4m Milankovi\'{c} telescope of AS Vidojevica
\end{table}

\section{Observations and data reduction}
\label{obs}

Our first observations of (300) Geraldina were conducted over two consecutive nights, 10/11 and 11/12 January 2002, using the 2m telescope at BNAO Rozhen. In Tab.\,\ref{table_physpar} the physical parameters such as diameter, absolute magnitude, albedo, and orbital parameters of (300) Geraldina are shown (as can be found in the JPL Small-Body Database Browser and Asteroid Light Curve Database\footnote{https://ssd.jpl.nasa.gov/tools/sbdb\_lookup.html}).

\begin{table}
\begin{center}
\caption{Asteroid parameters}
\label{table_physpar}
\begin{tabular}{lcccccc}
\hline
\hline
Asteroid	&	$D$ (km)	& $H$ (mag)	& Albedo	& $a$ (AU)	&	$e$	&	$i$ ($^{\circ}$)\\
\hline
(300) Geraldina	&	67.37	&	9.76	&	0.052	&	3.20	&	0.06	&	0.73\\
\hline
\hline
\end{tabular}
\end{center}
\end{table}

From the acquired photometric data, we derived—for the first time in the case of this asteroid—a synodic rotation period of 6.818 hours and amplitude of 0.317\,mag \citep{2002ESASP.500..505I}. \cite{2006MPBu...33...50L} from his observations in 2005 calculated a more precise rotational period of 6.842\,h.

The next observations at BNAO Rozhen were performed with the 50/70cm Schmidt telescope and with the 60cm Cassegrain telescope. Since 2017, Geraldina was observed at Astronomical Station Vidojevica with the 1.4m Milankovi\'{c} telescope and 60cm Cassegrain Nedeljkovi\'{c} telescope. The larger Milankovi\'{c} telescope was installed in 2016, and asteroids Geraldina and O'Steen were among the first targets observed with it, to determine their shapes \citep{2020IAUGA..30...39T}. During 2002 and 2003, photometric observations were done in the standard Johnson–Cousins V band and all others in the R band.

\begin{table}[H]
\small
\begin{center}
\caption{Aspect data  for the asteroid (300) Geraldina}
\label{aspect_data}
\begin{tabular}{lccccccl}
\hline
\hline
Date     &   $r$    &  $\Delta$  & $\alpha$ &   $PLON$ & $PLAT$ & Filter & Note\\
(UT)     &   (AU)    &  (AU)  & ($^{\circ}$) &   ($^{\circ}$) & ($^{\circ}$) &  & \\
\hline
2002 Jan 11.01	&	3.368	&	2.406	&	9.39	&	138.74	&	0.88	&	V	&	2mRCC\\
2002 Jan 12.01	&	3.276	&	2.399	&	9.09	&	138.76	&	0.88	&	V	&	2mRCC\\
2003 Apr 28.92	&	3.368	&	2.381	&	4.02	&	206.59	&	0.22	&	V	&	5070\\
2003 Apr 29.96	&	3.367	&	2.384	&	4.37	&	206.57	&	0.22	&	V	&	5070\\
2003 Apr 30.90	&	3.367	&	2.387	&	4.68	&	206.56	&	0.21	&	V	&	5070\\
2008 Jan 09.08	&	3.319	&	2.656	&	14.06	&	155.98	&	0.81	&	R	&	5070\\
2008 Jan 09.92	&	3.319	&	2.644	&	13.87	&	156.04	&	0.81	&	R	&	5070\\
2008 Jan 11.02	&	3.320	&	2.633	&	13.68	&	156.11	&	0.81	&	R	&	5070\\
2009 Apr 24.05	&	3.353	&	2.363	&	3.66	&	224.28	&	0.00	&	R	&	5070\\
2009 Apr 25.06	&	3.353	&	2.360	&	3.31	&	224.27	&	0.00	&	R	&	5070\\
2010 Jul 13.94	&	3.161	&	2.150	&	2.30	&	297.25	&	-0.88	&	R	&	5070\\
2010 Jul 14.98	&	3.161	&	2.148	&	1.94	&	297.24	&	-0.88	&	R	&	5070\\
2010 Aug 17.83	&	3.144	&	2.241	&	9.90	&	297.34	&	-0.88	&	R	&	5070\\
2011 Oct 30.96	&	3.037	&	2.078	&	5.90	&	21.62	&	-0.29	&	R	&	60CAS\\
2011 Nov 29.81	&	3.040	&	2.327	&	14.70	&	22.93	&	-0.19	&	R	&	60CAS\\
2011 Nov 30.81	&	3.040	&	2.338	&	14.91	&	23.01	&	-0.18	&	R	&	60CAS\\
2012 Nov 12.08	&	3.178	&	2.566	&	15.66	&	102.12	&	0.6	&	R	&	60CAS\\
2012 Nov 13.95	&	3.179	&	2.545	&	15.33	&	102.28	&	0.65	&	R	&	60CAS\\
2012 Dec 13.04	&	3.195	&	2.283	&	7.85	&	103.58	&	0.74	&	R	&	60CAS\\
2013 Jan 08.85	&	3.210	&	2.230	&	1.90	&	103.33	&	0.79	&	R	&	60CAS\\
2014 Mar 30.94	&	3.380	&	2.423	&	5.70	&	173.15	&	0.64	&	R	&	60CAS\\
2017 Oct 20.08	&	3.050	&	2.101	&	6.79	&	44.44	&	-0.03	&	R	&	EQ60\\
2018 Dec 04.01	&	3.248	&	2.612	&	14.87	&	121.75	&	0.79	&	R	&	60CAS\\
2018 Dec 07.11	&	3.250	&	2.576	&	14.29	&	121.98	&	0.79	&	R	&	60CAS\\
2018 Dec 12.97	&	3.252	&	2.513	&	13.06	&	122.34	&	0.81	&	R	&	AZ1400\\
2020 Feb 28.02	&	3.387	&	2.546	&	10.21	&	190.79	&	0.51	&	R	&	60CAS\\
2020 Apr 18.96	&	3.386	&	2.434	&	6.36	&	190.33	&	0.43	&	R	&	2mRCC\\
2023 Oct 07.04	&	3.088	&	2.462	&	16.25	&	64.82	&	0.20	&	R	&	EQ60\\
2023 Oct 15.00	&	3.092	&	2.374	&	14.70	&	65.52	&	0.22	&	R	&	60CAS\\
2025 Mar 05.90	&	3.350	&	2.449	&	8.33	&	140.08	&	0.84	&	R	&	AZ1400\\
\hline
\hline
\end{tabular}
\end{center}
Notes: In the last column, the telescope abbreviations are the same, as in Tab.\,\ref{table_instruments}
\end{table}

Geraldina was observed 30 nights during 13 oppositions over more than two decades from 2002 to 2025. In Tab.\,\ref{aspect_data}, the aspect data of (300) Geraldina during our observations are reported. The first column is the date of the observation referring to the mid–time of the observed lightcurve.  In the other columns are as follows: asteroid distance from the Sun ($r$), distance from the Earth ($\Delta$), Phase Angle $(\alpha)$ (the Sun-Asteroid-Observer angle), Phase Angle Bisector Longitude ($PLON$) and Phase Angle Bisector  Latitude ($PLAT$) of the asteroid referred to the time in the first column, the broadband filter in which the observations were carried. The last column is a note on the telescope used to acquire images, and an explanation of the abbreviations can be found at the bottom of Tab.\,\ref{table_instruments}.

\begin{figure}[hb]
\vspace*{2.mm}
\centerline{\includegraphics[width=0.99\textwidth,clip=]{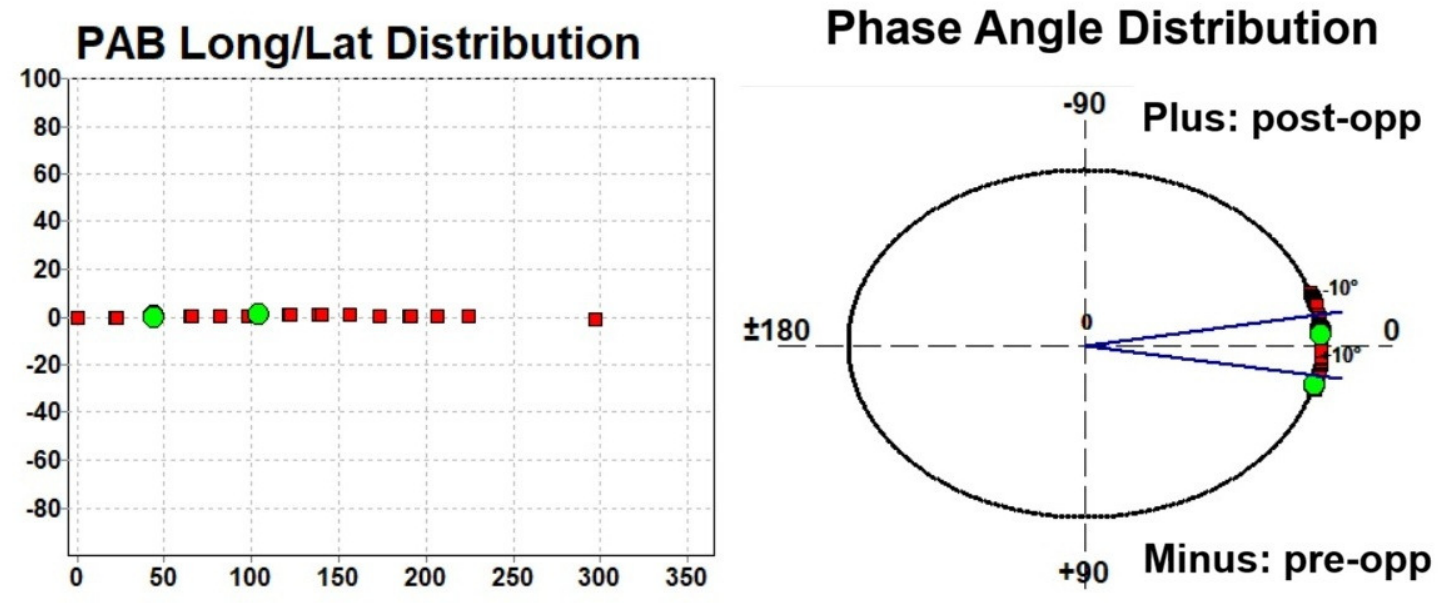}}
\vspace*{2.mm}
\caption{Left: Distribution of PAB longitude vs PAB latitude; Right: Distribution of the phase angle. On each of the two plots, the red and green points represent the dense and GAIA DR3 sparse observations, respectively.}
\label{PABdenseGaia}
\end{figure}

From the Asteroid Light Curve Data Exchange Format (ALCDEF)\footnote{https://alcdef.org/} database \citep{2018DPS....5041703S}, for (300) Geraldina, we collected dense photometric data from two LCs from \cite{2015AJ....150...75W} and six LCs from  \cite{2018MPBu...45..199P}. Sparse data correspond to 37 photometric points from 11 August 2014 to 02 January 2017 from GAIA Data Release 3 (DR3) \citep{2023A&A...674A...1G, 2023A&A...674A..12T}. Our dense photometric data provide broad coverage of the phase-angle bisector (PAB) longitude distribution, while the Gaia sparse data largely overlap with already sampled regions of the PAB longitude range (Fig.\,\ref{PABdenseGaia}).

The data were reduced using standard bias and flat-field corrections, followed by aperture photometry using CCDPHOT by \cite{1996ASPC..101..135B}.

\section{Results}

\subsection{Lightcurve analysis}

\begin{figure}[h]
\centerline{\includegraphics[width=0.92\textwidth,clip=]{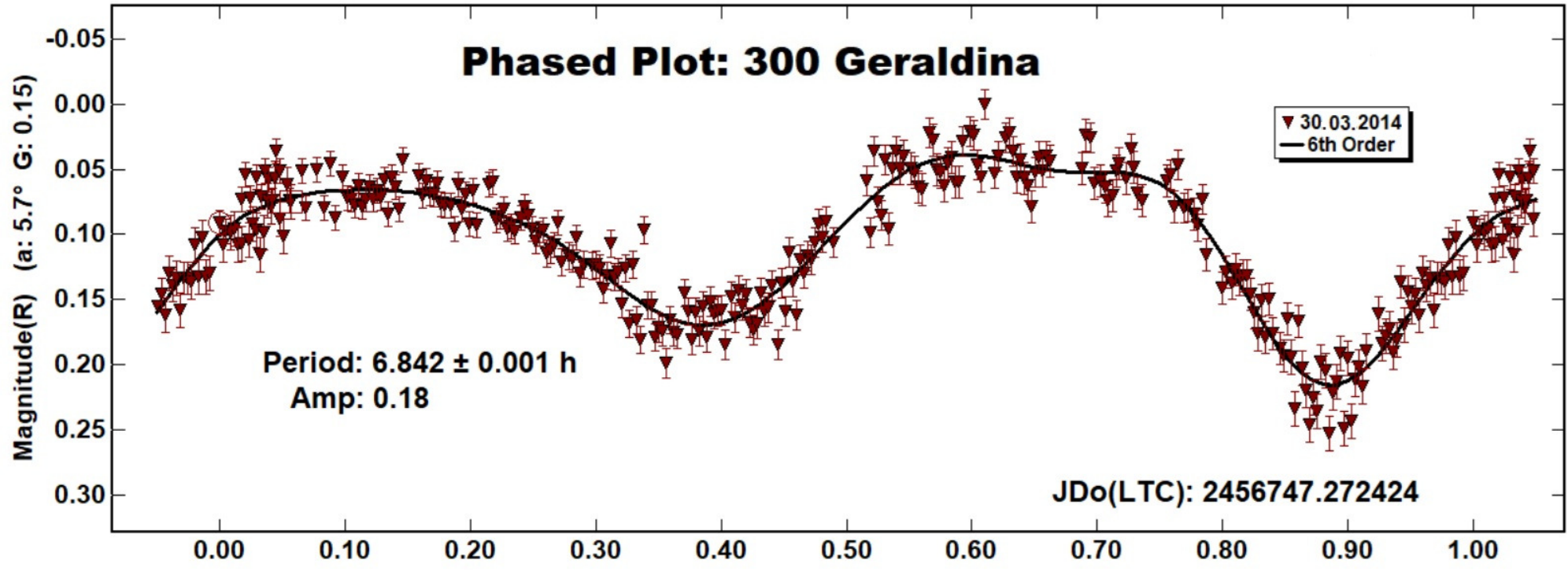}}
\caption{The single-night lightcurve of (300) Geraldina was obtained from observations carried out on 30 March 2014. Fourier analysis of the lightcurve, constructed using the accepted rotational period of 6.842\,h and a 6\textsuperscript{th}order Fourier fit, yielded an amplitude of 0.18\,mag.}
\label{fig02}
\end{figure}

\begin{figure}[h]
\centerline{\includegraphics[width=0.92\textwidth,clip=]{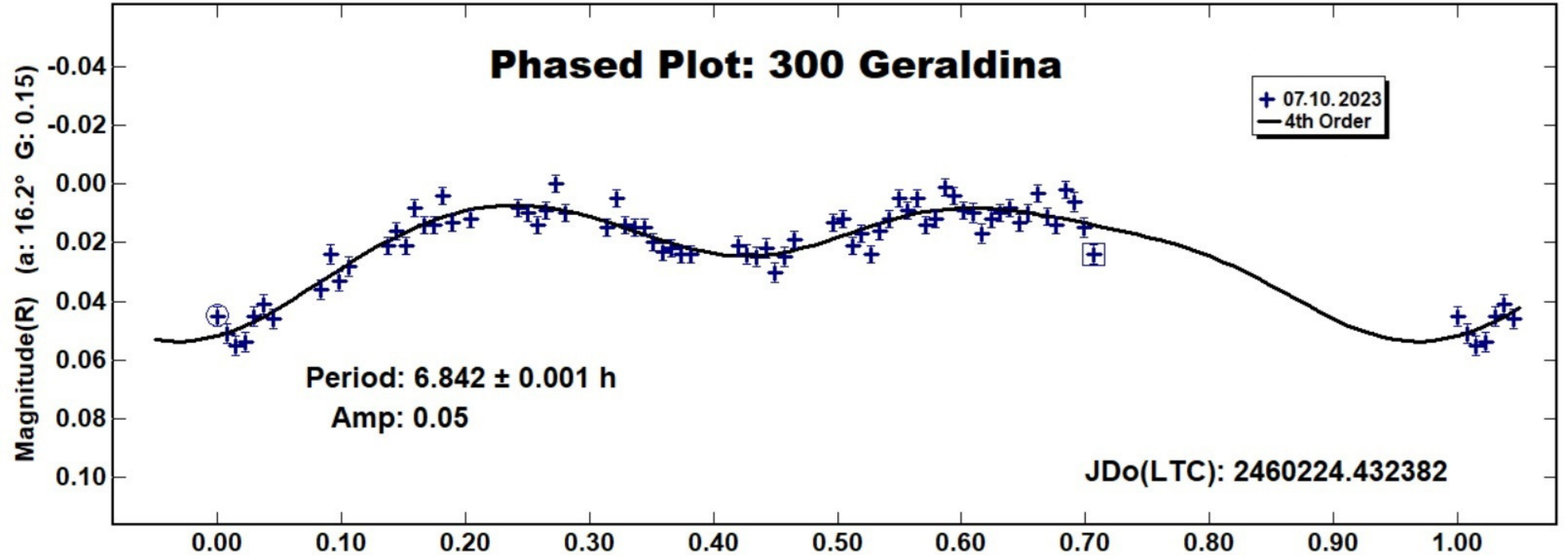}}
\caption{The single-night lightcurve of (300) Geraldina was obtained from observations carried out on 7 October 2023. Fourier analysis of the lightcurve, constructed using the accepted rotational period of 6.842\,h and a 4\textsuperscript{th}order Fourier fit, yielded an amplitude of 0.05\,mag.}
\label{fig03}
\end{figure}

\begin{figure}[h]
\centerline{\includegraphics[width=0.92\textwidth,clip=]{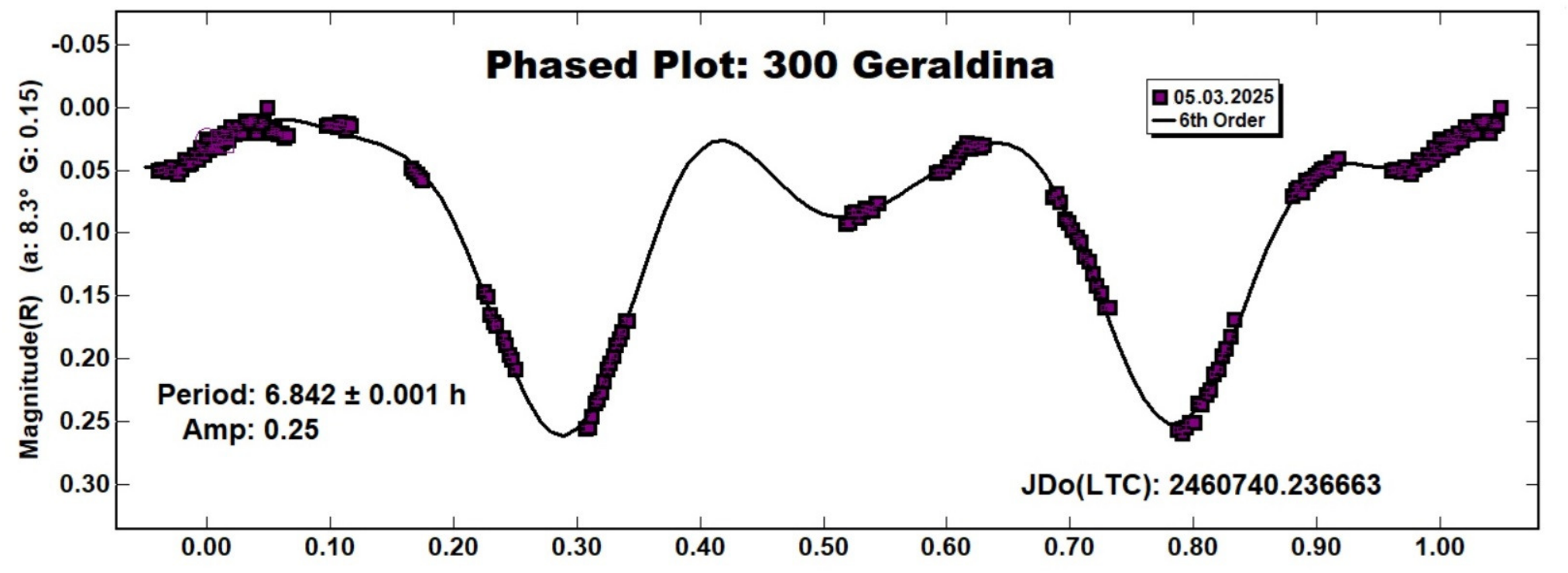}}
\caption{The single-night lightcurve of (300) Geraldina was obtained from observations carried out on 5 March 2025. Fourier analysis of the lightcurve, constructed using the accepted rotational period of 6.842\,h and a 6\textsuperscript{th}order Fourier fit, yielded an amplitude of 0.25\,mag.}
\label{fig04}
\end{figure}

Lightcurves obtained during three different oppositions are presented in Fig.\,\ref{fig02}, Fig.\,\ref{fig03}, and Fig.\,\ref{fig04}, illustrating how both the shape of the lightcurve and the amplitude change between apparitions.

\begin{figure}
\centerline{\includegraphics[width=0.9\textwidth,clip=]{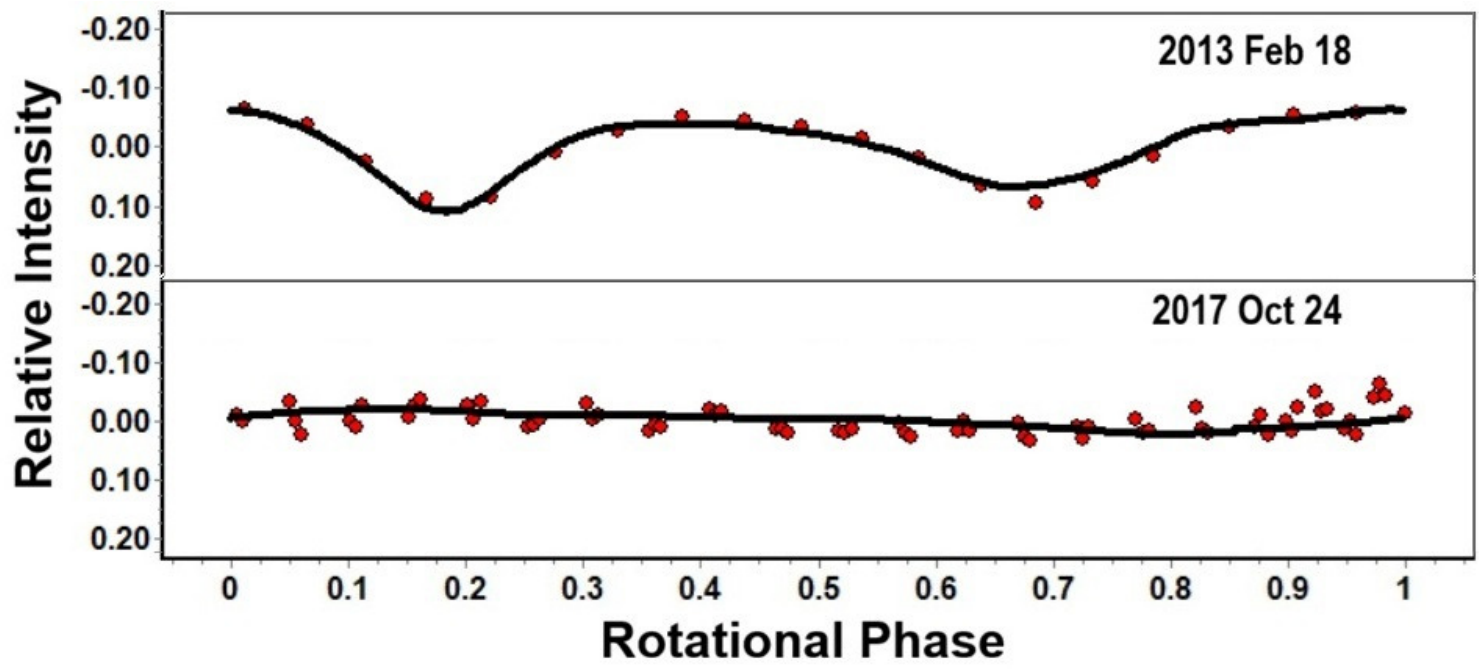}}
\caption{The observed lightcurves derived from available published data from ALCDEF from 2013 and from 2017 (red points) for selected nights are shown superimposed on the synthetic lightcurves generated by the shape–spin model obtained through a combined inversion of all dense datasets and the GAIA DR3 sparse photometry.}
\label{fig05}
\end{figure}

\begin{figure}[pt]
\centerline{\includegraphics[width=0.64\textwidth,clip=]{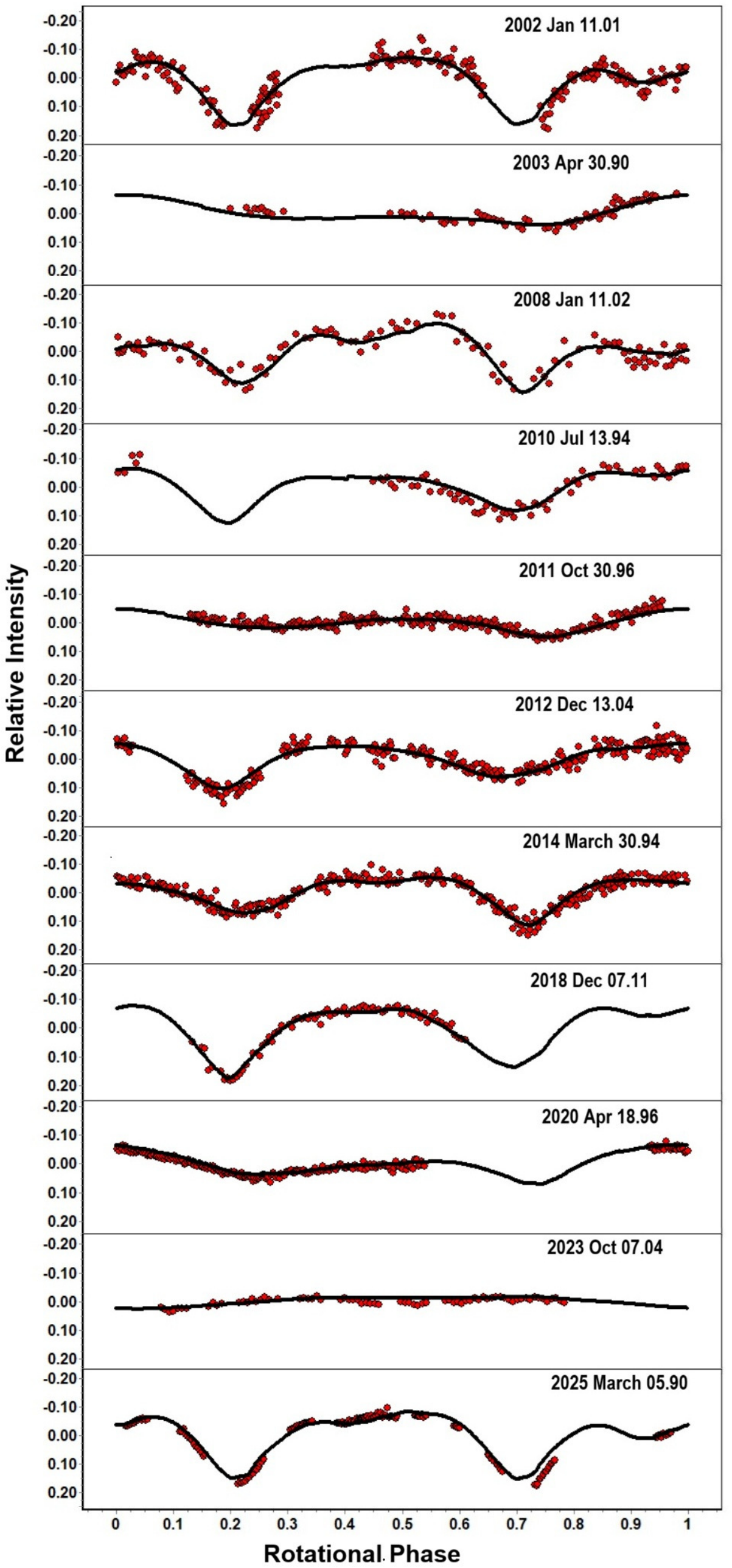}}
\caption{The observed lightcurves derived from our dense photometric measurements (red points) for selected nights are shown superimposed on the synthetic lightcurves generated by the shape–spin model obtained through a combined inversion of all dense datasets and the GAIA DR3 sparse photometry.}
\label{fig06}
\end{figure}

For the determination of a shape model, we used the lightcurve inversion method \citep{2001Icar..153...24K, 2001Icar..153...37K} and the LCInvert programme, part of the MPO Software. Considering the high accuracy of the DR3 photometric data, we took the weighting factor to be set to 1.0 for both sparse and for dense photometric data. The rotational period was determined using a light-curve inversion method that fits a simple 3D shape model to the data, instead of standard Fourier techniques. The best period was selected by minimizing the $\chi^2$  value computed for the modeled and observed light curves. The analysis was carried out with the DAMIT software package \citep{2001Icar..153...37K, 2010A&A...513A..46D}. Trial periods were scanned over a wide range using the \textit{period\_scan} routine, testing six initial pole solutions for each trial period,  to ensure that the global $\chi^2$ minimum was found.

\begin{figure}[pt!]
\centerline{\includegraphics[width=0.53\textwidth,clip=]{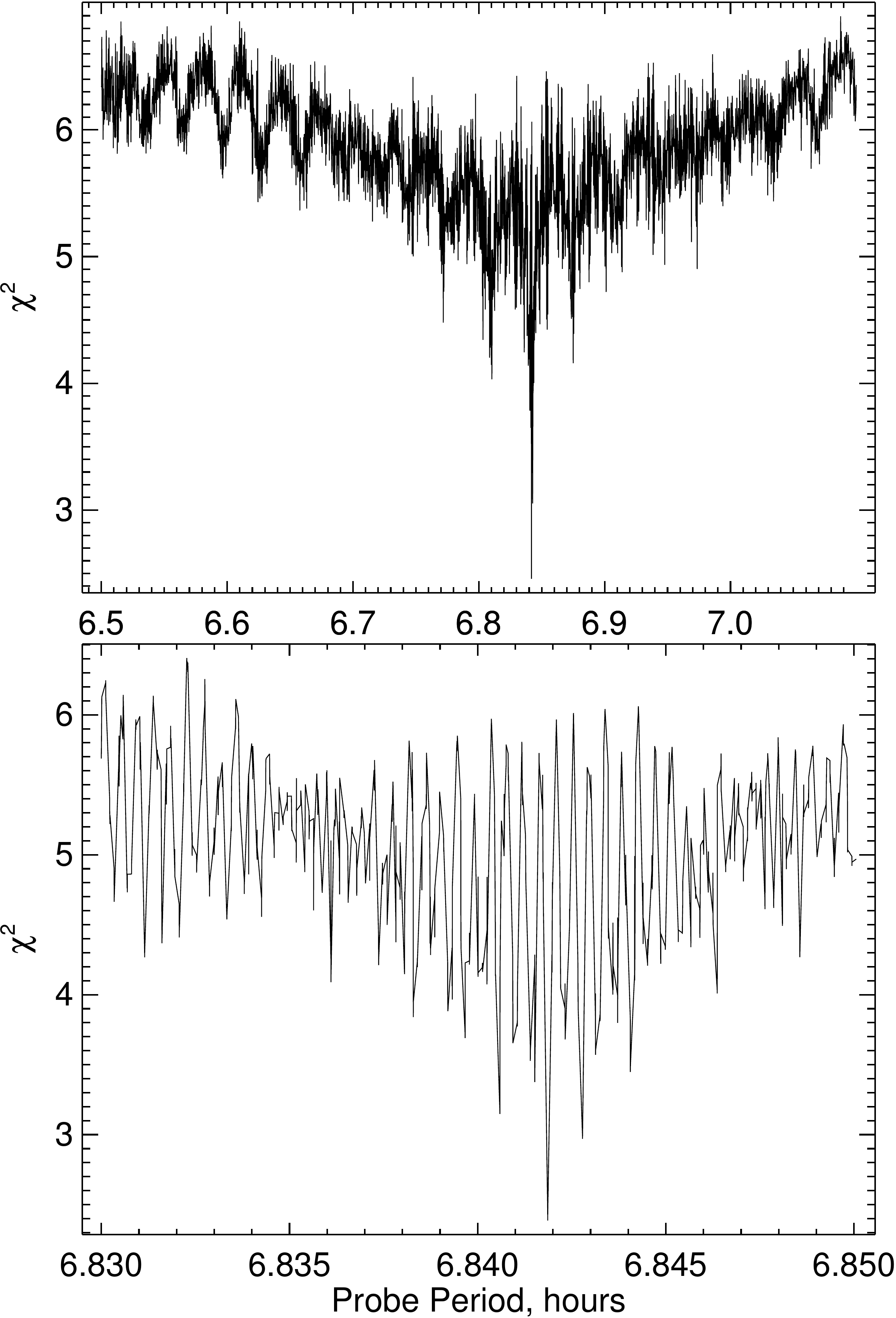}}
\caption{$\chi^2$ vs probe period plot used for a period search (top panel) and the same plot but zoomed around the best solution with the lowest $\chi^2$ value (bottom panel)}
\label{300period}
\end{figure}
\begin{figure}[pb!]
\centerline{\includegraphics[width=0.765\textwidth,clip=]{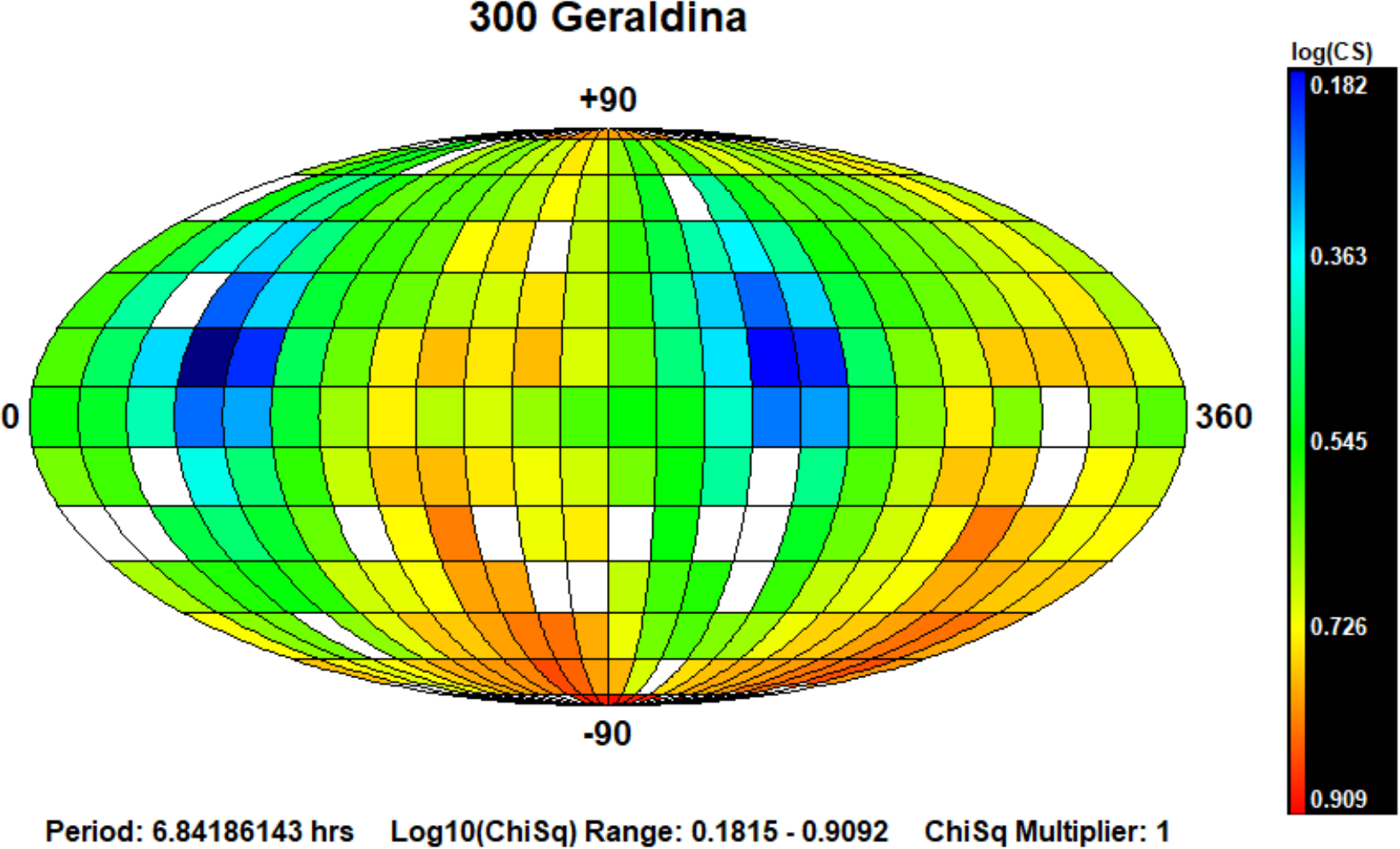}}
\caption{Pole search distribution. The dark blue regions have the smallest value of $\chi^2$.}
\label{polemap}
\end{figure}

We made the initial sidereal period search around 6.8\,h within the interval of $\pm0.3$\,h (see Fig.\,\ref{300period} top panel). Several solutions with 10\% of the lowest $\pm2$ value were found, but a narrower period search found that the most likely period is 6.84186100\,h (see Fig.\,\ref{300period} bottom panel.)

The pole search produced 312 discrete and fixed pole positions distributed over the unit sphere with 15$^{\circ}$ steps in ecliptic latitude and longitude, varying also the most likely input period. The resulting solutions were used to construct $\chi^2$ maps (Fig.\,\ref{polemap}), which allowed the identification of low-$\chi^2$ regions and the selection of the best-fitting pole. As it is expected for asteroids with orbits close to the ecliptic plane, we encounter two symmetric pole solutions with identical $\beta$ and $\lambda$ $\pm$ 180$^{\circ}$. The preliminary ecliptic coordinates for the pole are $\lambda = 45 \pm 7.5 ^{\circ}$, $\beta = 15 \pm 7.5 ^{\circ}$ and for the mirrored pole solution are $\lambda = 225 \pm 7.5 ^{\circ}$, $\beta = 15 \pm 7.5 ^{\circ}$ which suggest prograde sense of asteroid rotation. The finest search centered on these roughly obtained pole positions gave us more precisely determined solutions, presented in Tab.\,\ref{table_model}.

\begin{table}
\begin{center}
\caption{Parameters of the model of the asteroid (300) Geraldina}
\label{table_model}
\begin{tabular}{ccccccc}
\hline
\hline
Sideral  		& Sense of 					& \multicolumn{2}{c}{Pole}&	$a/b$	&	$a/c$	&	$b/c$\\
period (h)	&	rotation					&	$\lambda$ ($^{\circ}$) & $\beta$ ($^{\circ}$)	&				&			&			\\
\hline
6.8418615	&	P	&	$52.6 \pm 5.0$		&	$9.5 \pm 5.0$	&	1.11	&	1.29	&	1.16	\\
6.8418615	&	P	&	$232.6 \pm 5.0$	&	$9.1 \pm 5.0$	&	1.11	&	1.29	&	1.16	\\
\hline
\hline
\end{tabular}
\end{center}
\end{table}

Assuming the asteroid as a triaxial ellipsoid, rotating around the shortest axis $c = 1\quad(a > b > c)$, we derived relative shape dimensions $a/c = 1.29$.  The Tab.\,\ref{table_model} gives the sidereal rotational period, sense of rotation (P for prograde), ecliptic coordinates $\lambda$ and $\beta$ of the two mirror solutions, and rough relative shape dimensions.

\begin{figure}
\centerline{\includegraphics[width=0.8\textwidth,clip=]{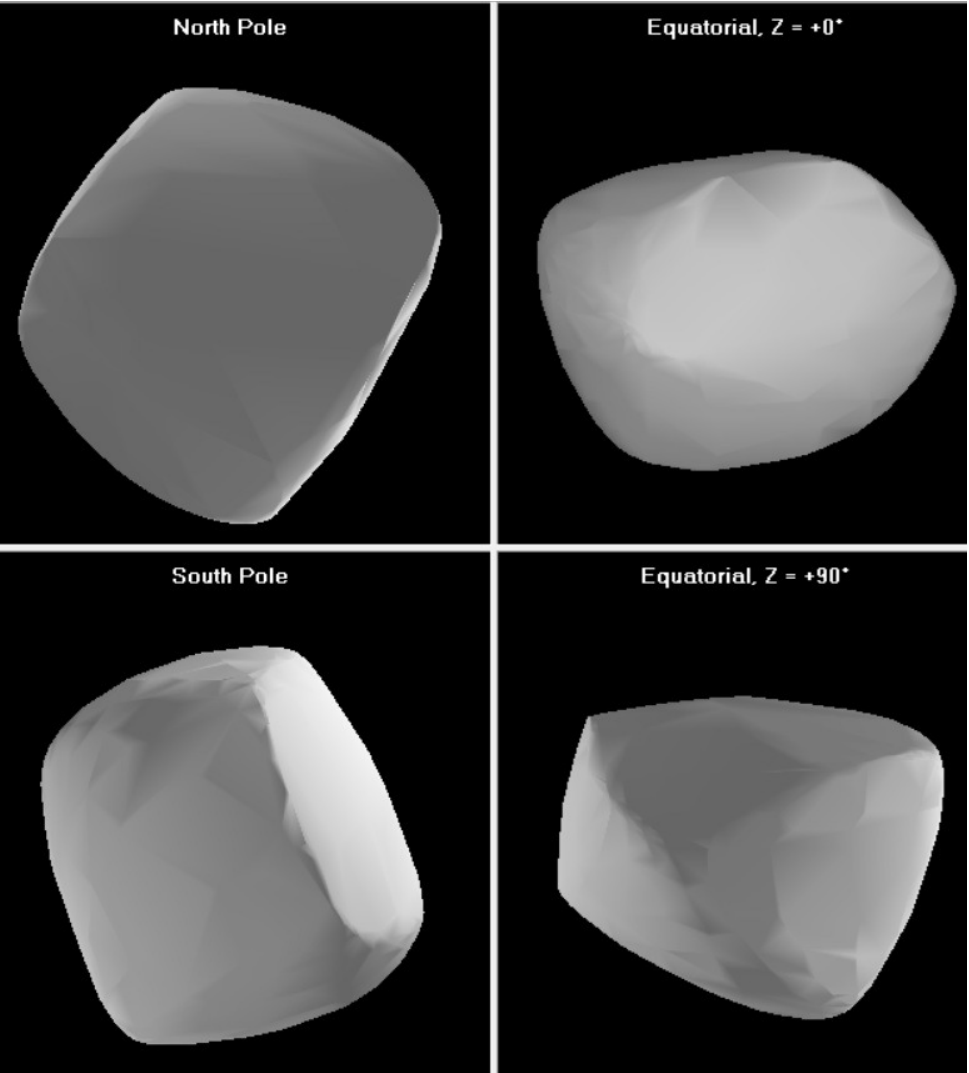}}
\caption{The three-dimensional low-resolution convex shape model of the asteroid (300) Geraldina is shown. The left panels present views of the model from the north and south rotational poles, illustrating the overall polar morphology. The right panels display equatorial perspectives separated by $90^{\circ}$ in rotational phase, highlighting variations in the projected shape and allowing assessment of the object's elongation and symmetry during rotation.}
\label{model45}
\end{figure}

Using the maximum observed lightcurve amplitude reported by \cite{2002ESASP.500..505I}, and applying the empirical relation for C-type asteroids proposed by \cite{1990A&A...231..548Z}, we calculated the ratio of the maximum to minimum projected reflecting surface areas during the asteroid’s rotation, as done in \cite{2021SerAJ.202...39V} and \cite{2024CoSka..54d..57B}, obtaining $a/c = 1.39$. None of the composite lightcurves raises a suspicion that this might be a binary asteroid as suggested by \cite{2024MPBu...51..213R}. The minima have approximately the same depths and are separated by half a phase. If (300) Geraldina were a binary asteroid, its lightcurve should have a more irregular shape, as in the case of the suspected binary (12499) 1998 FR47 \citep{2024CoSka..54d..57B}.

\subsection{Taxonomy from Gaia spectroscopy}

The Gaia mission has been observing Solar System objects since the beginning of its operation, and the mean reflectance spectra of selected asteroids are included in its 3rd data release (DR3). As the asteroid (300) Geraldina is among the asteroids with reflectance spectra in DR3, we will use it in order to obtain its taxonomy and composition, relying on the new probabilistic approach by \cite{2022A&A...665A..26M}.

\begin{figure}
\centerline{\includegraphics[width=0.99\textwidth,clip=]{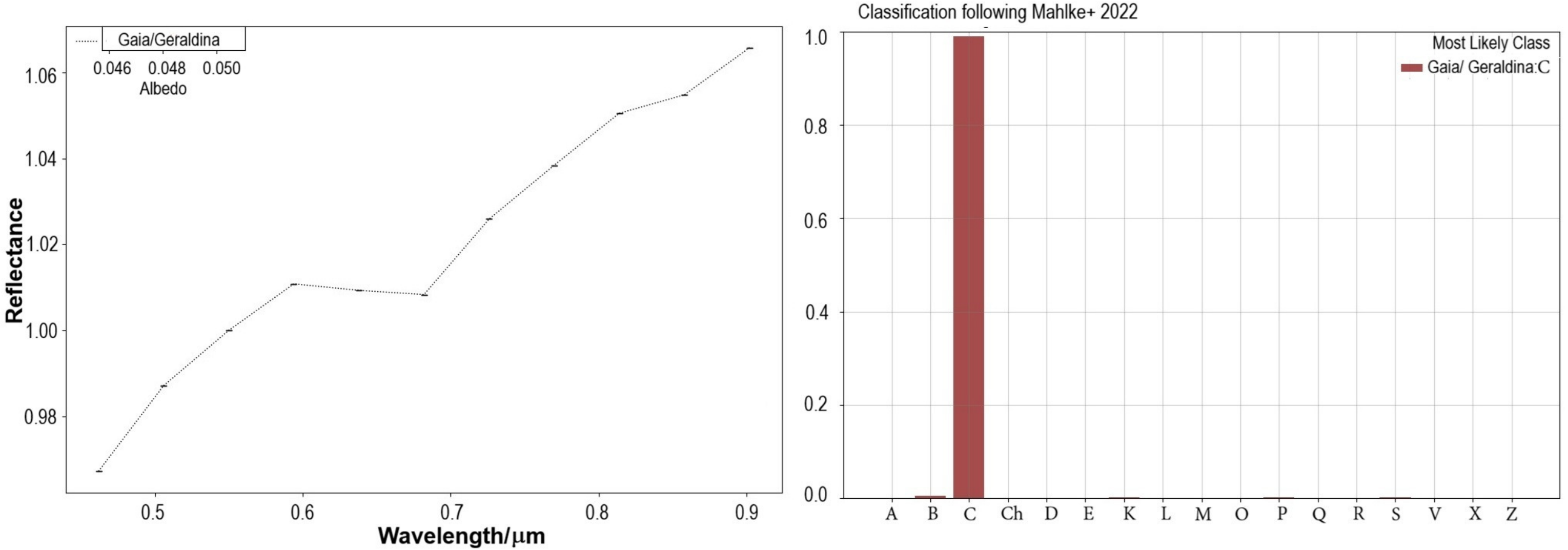}}
\caption{Gaia DR3 reflectance spectrum of (300) Geraldina and its taxonomical classification according to \cite{2022A&A...665A..26M} (see text for details).}
\label{DR3_sp}
\end{figure}

The results are shown in Fig.\,\ref{DR3_sp}. The left panel shows the Gaia DR3 reflectance spectra of the object, and the right panel presents the possible taxonomy classes with their probabilities. In particular, the asteroid (300) Geraldina has a 99\% probability of being a C-class object and 1\% for a B-class. Both of these solutions fall into classes that are primitive carbonaceous. The previous taxonomy determination shows the same classification \citep{2010A&A...510A..43C}. This data set contains Sloan Digital Sky Survey (SDSS) asteroid photometric observations classified according to the SDSS-based Asteroid Taxonomy, as developed by \citeauthor{2010A&A...510A..43C}. Taking into account its reported very low albedo - 0.0331 \citep{2010A&A...510A..43C} may also lead to the idea that this object could be an "ancient asteroid"!

\subsection{Dynamical properties of the asteroid (300) Geraldina}

Dynamical properties of the asteroid (300) Geraldina are studied in two ways. First, we integrated the asteroid over the time span of 100\,Myrs with the Orbit9 software\footnote{http://adams.dm.unipi.it/orbfit/}, incorporating all the planets in the Solar system model. One common way to observe whether an asteroid has undergone dynamical evolution is to track changes in its semi-major axis.

\begin{figure}
\centerline{\includegraphics[width=0.99\textwidth,clip=]{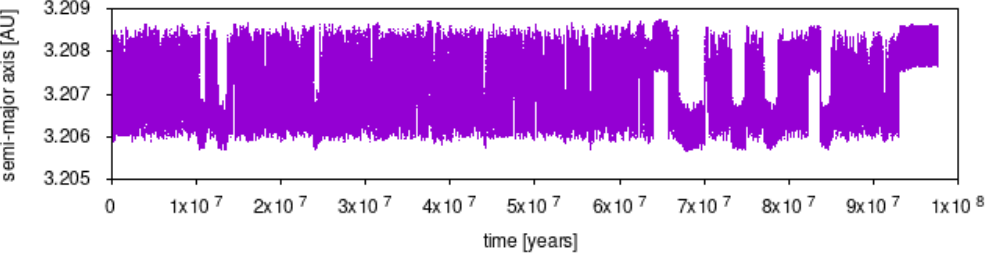}}
\caption{The stability of the asteroid (300) Geraldina over 100\,Myrs shown through the evolution of its semi-major axis $a$. The asteroid remains stable, exhibiting only local oscillation in $a$, with several resonant captures (visible as intervals with oscillations of smaller amplitudes) that stabilize the asteroid even more. These captures became more frequent and lasted longer after 67\,Myrs.}
\label{figstable}
\end{figure}

Our results (see Fig.\,\ref{figstable}) show that (300) Geraldina is a stable asteroid. Its semi-major axis $(a)$ exhibits only local oscillations ranging between a=[3.206, 3.209]\,AU, along with several sporadic resonant captures that stabilize the asteroid even more. In Fig.\,\ref{figstable}, these resonant episodes are visible as time intervals where the oscillation amplitudes are smaller. Before approximately $67$\,Myrs, the longest stays in the resonance are on the order of 100\,Kyrs.  After $\sim 67$\,Myrs, they become longer, with the final and longest resonant episode lasting from 90.3\,Myr to the end of the integration.

\begin{figure}
\centerline{\includegraphics[width=0.99\textwidth,clip=]{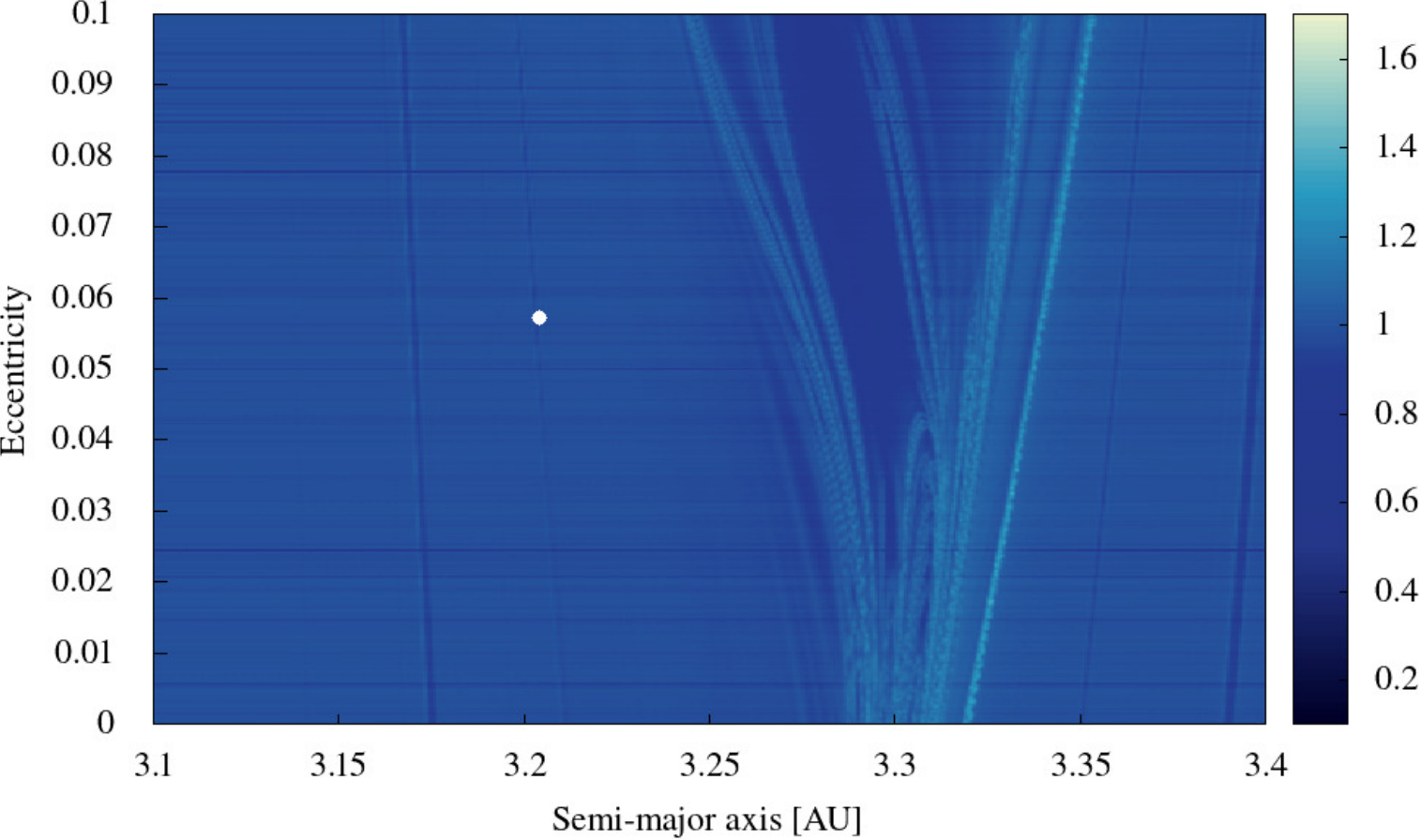}}
\caption{The FLI map showing the dynamical neighborhood of the asteroid (300) Geraldina (marked with a white dot). Lighter colors reveal instabilities, while darker shades of blue represent stable regions. The most dominant structure revealing instability appears around 3.3\,AU, and represents one of the strongest resonances in the Solar system - the 2:1 MMR with Jupiter. Although relatively close, (300) Geraldina has not been captured by this resonance during 100\,Myrs. The only resonant behavior of the asteroid is observed within  a weak high-order MMR, visible on the map as a thin vertical line at about $a \sim 3.21$\,AU.}
\label{FLImap}
\end{figure}

We also investigated the dynamical environment of the asteroid. Fig.\,\ref{FLImap} shows an Fast Lyapunov Indicator - FLI map \citep{1997P&SS...45..881F, 1997CeMDA..67...41F, 2024CoSka..54d..57B, 2017MNRAS.465.4441T}, illustrating the stability properties within a portion of the orbital plane surrounding the asteroid, defined by $[a \times e] =$\linebreak$ [3.1 \mathrm{AU}, 3.4 \mathrm{AU}] \times [0, 0.1]$. The position of (300) Geraldina at $(a,e) =$\linebreak$ (3.204 \mathrm{AU}, 0.057)$ is indicated by a white dot, while the remaining orbital elements—the inclination $i$, longitude of the ascending node $\Omega$, argument of perihelion $\omega$, and mean anomaly $M$—are set equal to those of Geraldina at epoch JD 2458800.5, obtained from NASA’s JPL Horizons system\footnote{https://ssd.jpl.nasa.gov/horizons.cgi}. This choice of orbital parameters provides a realistic representation of the dynamical environment around the asteroid. The map was computed using the Orbit9 integrator over a timespan of 5,000 years, adopting a full Solar System model.

Each point on Fig.\,\ref{FLImap} represents one orbit. Pixels colored in light blue represent orbits that have shown instabilities during 5000 years, while the darker shades show stable regions. The most dominant structure on the map is noticeable at 3.3\,AU. It represents the trace of one of the strongest resonances in the Solar system, the 2:1 mean motion resonance (MMR) with Jupiter \citep{Chrenko2015MNRAS.451.2399C, Broz2005dpps.conf..179B}. Although (300) Geraldina lies relatively close to this resonance, it has not been driven toward it during the 100\,Myr integration (see changes in $a$ in Fig.\,\ref{figstable}). Let us mention that the integration does not include the Yarkovsky effect, which could, in principle, cause an asteroid to drift into the resonance \citep{2005MNRAS.359.1437B}. However, large asteroids, such as Geraldina, are, in general, not significantly affected by the Yarkovsky force.

Several thin and pale vertical lines in Fig.\,\ref{FLImap} represent unidentified high-order MMRs. They are not strong enough to make any significant change in the orbital evolution of an asteroid. Geraldina appears to be on one such resonance. The changes of its semi-major axis shown in Fig.\,\ref{figstable} do reveal a resonant behavior, but as mentioned above, it was not strong enough to provide any significant shift of the asteroid. This weak MMR only contributed to the stabilization of the asteroid. Geraldina's dynamical stability, although not evaluated over a full 4 billion–year timescale, may provide additional evidence of its pristine nature.

\section{Conclusions}

We have presented a combined photometric, spectroscopic, and dynamical analysis of the main-belt asteroid (300) Geraldina, using dense ground-based light\-curves spanning more than two decades, complemented by available published data from ALCDEF and by sparse Gaia DR3 photometry. Application of the lightcurve inversion method yielded a well-constrained sidereal rotation period of 6.8418615\,h, a prograde sense of rotation, and two mirror pole solutions consistent with the asteroid’s low orbital inclination. The derived convex shape model indicates a moderately elongated body. Gaia DR3 reflectance spectroscopy classifies (300) Geraldina as a C-type asteroid, consistent with a primitive, carbonaceous composition. Together with its reported low albedo, this supports its classification as an ancient asteroid candidate.

Numerical integrations over 100\,Myr show that (300) Geraldina remains dynamically stable. Despite its proximity to the 2:1 mean-motion resonance with Jupiter, no resonance capture is observed, and the asteroid instead exhibits weak interactions with a high-order mean-motion resonance without much influence on its long-term orbital evolution.

Taken together, the physical, rotational, taxonomic, and dynamical properties of (300) Geraldina support the hypothesis that it is a primordial asteroid. These results make (300) Geraldina a valuable target for further studies of the primordial collisional and dynamical evolution of the asteroid belt.

\acknowledgements
This article is based on data collected with the telescopes of the Rozhen National Astronomical Observatory. The authors gratefully acknowledge observing grant support from the Institute of Astronomy and Rozhen National Astronomical Observatory, Bulgarian Academy of Science. The research that led to these results was partially carried out with the help of infrastructure purchased under the National Roadmap for Research Infrastructure, financially coordinated by the Ministry of Education and Science of Republic of Bulgaria. G.B. and A.K. acknowledge partial support by grant: K$\Pi$-06-H88/5 "Physical properties and chemical composition of asteroids and comets - a key to increasing our knowledge of the Solar System origin and evolution." by the Bulgarian Scientific Research Fund of the Ministry of Education and Science. The observations at AS Vidojevica and funding for N.T. were financed from the  Ministry of Science, Technological Development and Innovation of the Republic of Serbia, under the contract number 451-03-33/2026-03/200002. The authors are grateful to D. Athanasopoulos for valuable discussions and insights on ancient asteroids. The authors gratefully acknowledge Violeta Ivanova, Bonka Bilkina, and Vasko Umlenski for their contribution to the observational data collected at the National Astronomical Observatory Rozhen and used in this study.

\bibliography{Apostolovska_2026}
\end{document}